\documentclass{aa}
\usepackage[varg]{txfonts}

\usepackage{lscape}
\usepackage{graphicx}
\usepackage{colortbl}
\usepackage{amssymb}
\usepackage{amsmath}
\usepackage{natbib}
\bibpunct{(}{)}{;}{a}{}{,} 
\usepackage{times}
\usepackage{aas_macros}
\usepackage{threeparttable}
\usepackage{subfigure}
\usepackage{multirow}
\usepackage{float}

\usepackage{color}

\begin{document}

\title{Bulges and discs in the local Universe. Linking the galaxy structure to star formation activity}
\titlerunning{The galaxy morphology - star formation activity link}
\authorrunning{L. Morselli}
\author{L. Morselli\inst{1}
  \and P. Popesso \inst{1} 
  \and G. Erfanianfar \inst{1}
  \and A. Concas \inst{1}}
\institute{Excellence Cluster Universe, Boltzmannstr. 2, 85748 Garching bei M\"unchen, Germany}

\date{Received 26 July 2016 / Accepted ??}


\label{firstpage}

\abstract{We use a sample built on the SDSS DR7 catalogue and the bulge-disc decomposition of Simard et al. (2011) to study how the bulge and disc components contribute to the parent galaxy's star formation activity, by determining its position in the star formation rate (SFR) - stellar mass (M$_{\star}$) plane at 0.02$<z<$0.1. For this purpose, we use the bulge and disc colours as proxy for their SFRs. We study the mean galaxy bulge-total mass ratio (B/T) as a function of the residual from the MS ($\Delta_{MS}$) and find that the B/T-$\Delta_{MS}$ relation exhibits a parabola-like shape with the peak of the MS corresponding to the lowest B/Ts at any stellar mass. The lower and upper envelop of the MS are populated by galaxies with similar B/T, velocity dispersion and concentration ($R_{90}/R_{50}$) values. The mean values of such distributions indicate that the majority of the galaxies are characterised by classical bulges and not pseudo-bulges. Bulges above the MS are characterised by blue colours or, when red, by a high level of dust obscuration, thus indicating that in both cases they are actively star forming. When on the MS or below it, bulges are mostly red and dead. At stellar masses above $10^{10.5} $M$_{\odot}$, bulges on the MS or in the green valley tend to be significantly redder than their counterparts in the quiescence region, despite similar levels of dust obscuration. This could be explained with different age or metallicity content, suggesting different evolutionary paths for bulges on the MS and green valley with respect to those in the quiescence region. The disc color anti-correlates at any mass with the distance from the MS, getting redder when approaching the MS lower envelope and the quiescence region. The anti-correlation flattens as a function of the stellar mass, likely due to a higher level of dust obscuration in massive SF galaxies. We conclude that the position of a galaxy in the LogSFR-LogM$_{\star}$ plane depends on the star formation activity of its components: above the MS both bulge and disk are actively star forming. The nuclear activity is the first to be suppressed, moving the galaxies on the MS. Once the disk stops forming stars as well, the galaxy moves below the MS and eventually to the quiescence region. This is confirmed by a large fraction ($\sim45\%$) of passive galaxies with a secure two component morphology, coexisting with a population of pure spheroidals. Our findings are qualitatively in agreement with the compaction-depletion scenario, in which subsequent phases of gas inflow in the centre of a galaxy and depletion due to high star formation activity move the galaxy across the MS before the final quenching episode takes place. }

\keywords{galaxies: star formation - Galaxy: structure - Galaxy: evolution }

\maketitle

\section{Introduction}
The Main Sequence (MS) of star-forming galaxies (SFGs) is a linear relation between the star formation rate (SFR) of a galaxy and its stellar mass (M$_{\star}$). This relation is close to linear, with a small scatter (0.2-0.3 dex) over a wide stellar mass range (e.g., Brinchmann et al. 2004; Noeske et al. 2007a; Daddi et al. 2007; Noeske et al. 2007b; Elbaz et al. 2007; Salim et al. 2007; Whitaker et al. 2012; Speagle et al. 2014; Pannella et al. 2015). It has also been shown that, while the slope of the relation remains $\sim1$ up to $z\sim2$, the normalisation changes drastically with redshift, where the SFR of galaxies at $z\sim 2$ is almost 20 times larger than that of galaxies at $z\sim0$. (e.g., Schreiber et al. 2015). The physical processes that cause this decrease in SF activity are generally grouped under the term {\textit{quenching}}, a topic at the core of galaxy evolution studies.


Quenching works by prohibiting SF in a galaxy, by acting on the cold gas reservoir. Processes that  expel cold gas from galaxies such as galactic winds driven by supernovae and massive stars are efficient in dark matter (DM) haloes with  $M < 10^{12}$M$_{\odot}$ \citep[e.g.][]{1986ApJ...303...39D,2000MNRAS.317..697E}. Above this mass threshold, more powerful outflows are required to counter the deeper DM halo potential well. In such haloes, accreting central BHs can quench star formation by either heating the gas in the disk \citep[quasar mode;][]{2006ApJS..163....1H,2006ApJ...652..864H,2010MNRAS.404..180D,2010A&A...518L.155F,2012ARA&A..50..455F,2014A&A...562A..21C}, or mechanically removing the gas from the inner regions through powerful radio jets \citep[radio mode;][]{2006MNRAS.365...11C,2006MNRAS.370..645B,2006MNRAS.366..499D,2007ARA&A..45..117M,2008MNRAS.390.1399B,2013MNRAS.436.3031V,2005Natur.433..604D,2007ApJ...658...65C,2009Natur.460..213C}. Besides ejecting the gas, the effects of quenching can also be caused by a lack of cold gas supply. In DM halos with $M > 10^{12}$ M$_{\odot}$,  the transition from cold to hot accretion \citep{2006MNRAS.368....2D} would prohibit further SF \citep{2014ApJ...789L..21C}. 
On the other hand, \cite{2009ApJ...707..250M}, propose morphological quenching where the growth of central mass concentration, i.e., a massive bulge, could simply stabilise a gas disc against fragmentation, thus preventing SF without affecting the cold gas reservoir in the galaxy. 
It has also been proposed that quenching could be a combination of all these processes. For example, using their cosmological zoom-in simulations, \cite{2015MNRAS.450.2327Z} show that quenching of high redshift SFGs is preceded by a compaction phase that creates the so called blue nuggets: SFGs morphologically similar to quiescent ones. This compaction phase is caused by strong inflows to the centre due to minor mergers, and/or counter--rotating gas and violent disc instabilities. The availability of dense cold gas in the centre results in a high SF and consequent stellar/supernova and/or AGN feedback causing quenching. \cite{2015Sci...348..314T} propose that episodes of compaction, quenching and replenishment are responsible for the position of a galaxy in a $\pm 0.3$ dex region around the MS. The final quenching episode results in a passive galaxy when the replenishment time is longer than the depletion time, typical of massive haloes, or low redshifts. 

There is increasing evidence in support of this picture from an observations. At $z\gtrsim2$, SFGs with the highest central gas densities are remarkably compact and have high S\'ersic indices ($n$) and spheroidal morphologies \citep{2011ApJ...742...96W,2014ApJ...791...52B,2013ApJ...766...15P,2013ApJ...768...92S,2013AAS...22131307W,2016ApJ...828...27N}. These galaxies resemble the quiescent population at the same redshift but are radically different from other SFGs that have irregular and clumpy appearances (\citealt{2004ApJ...604L..21E}; \citealt{2008ApJ...687...59G}; \citealt{2015ApJ...800...39G}). 
At low redshift, less massive SFGs are mainly pure disks, while more massive SFGs are characterised by a bulge+disc structure (\citealt{2012ApJ...753..114W}; \citealt{2012MNRAS.427.1666B}; \citealt{2014MNRAS.444.1660B}; \citealt{2014ApJ...788...11L}). \citet{2011ApJ...742...96W} analyse the dependence of galaxy structure (size and S\'{e}rsic index) on the position of the galaxies with respect to the MS ridge at $z=0-2$. They find that the S\'{e}rsic index tends to be roughly the same, $n\sim1$, within $3\sigma$ from the MS, i.e., there is no significant gradient of $n$ across the MS. They observe an increase of $n$ only in the starburst region. While Cheung et al. (2012) confirm that the S\'ersic index  most sharply discriminates between the red sequence and the blue cloud, they also find a large number of blue outliers ($\approx40\%$) in their $n>2.3$ galaxy sample at $z\sim 0.65$. Concurrently, using a sample of local galaxies \cite{Guo:2015kp} report that that bulges/bars are responsible for the large dispersion of the sSFR, but only for massive SFGs.  These recent results imply that a significant fraction of SFGs must have a bulge+disk morphology. 
Under the assumption that the star formation activity is confined in the disc, \cite{2014ApJ...785L..36A} show that defining the sSFR (specific SFR) with respect to the disc mass rather than the total galaxy mass decreases (and sometimes erases) the dependence of sSFR on stellar mass. This suggests that the slope of the sSFR - M$_{\star}$ relation reflects the increase of the bulge prominence with stellar mass, and also points towards the importance of treating galaxies as multicomponent systems.

The aim of this paper is to study the relation between galaxies structural parameters and location with respect to the MS in the local Universe. The choice of performing this study on a local galaxy sample is dictated by the fact that the Sloan Digital Sky Survey \citep[SDSS;][]{2002AJ....124.1810S} provides the required statistics to dissect the MS brick by brick to study with high accuracy the nature of its scatter as a function of the galaxy morphology. In addition, the availability of accurate morphological classification and bulge/disc decomposition (Simard et al. 2011) of the SDSS galaxies allows to study the role and the interconnection of the individual galaxy components. This is the first paper of a series, which follow the evolution of the relation of the galaxy structural parameter and the scatter across the MS in the $0-1.2$ redshift window. 

The paper is organised as follows. In Section 1 we describe our dataset. In Section 2 we analyse the reliability of the morphological classification and of the colours of the bulge and disc in individual galaxies. In Section 3 we present our results and in Section 4 we present our discussion and conclusions. Throughout the study, the following cosmology is assumed: $H_0=70.0\ km\ s^{-1}Mpc^{-1}$, $\Omega_m$=0.3 and $\Omega_{\Lambda}$=0.7. 

\section{Dataset}

In this paper we make use of the catalogues derived from the SDSS-DR7 database \citep{2009ApJS..182..543A} and in particular the following quantities: star formation rates, stellar masses, stellar velocity dispersion, and properties derived  from bulge/disc decomposition of galaxies. We will now briefly summarise how these were derived and refer the reader to the original papers for details.

\subsection{SFRs, stellar masses and velocity dispersion}

SFRs and M$_{\star}$ are taken from the MPA/JHU DR7 catalogues \citep{2003MNRAS.341...54K,2004MNRAS.351.1151B,2007ApJS..173..267S}.  In particular, we use the total, aperture corrected, stellar mass computed from the total \textit{ModelMag} photometry, less sensitive to emission line contamination. SFRs come from emission line modelling using the grid of $\sim$2x$10^5$ models of \cite{2001MNRAS.323..887C} where dust corrections are mainly based on the H$\alpha$/H$\beta$ ratio. 
The stellar aperture velocity dispersion $\sigma_{ap}$ (computed from stars), and mean S/N per pixel are taken from the MPA/JHU \textit{gal\_info} catalogue, while values of the H$\alpha$ and H$\beta$ fluxes and their S/N ratios are form the MPA/JHU \textit{gal\_line} catalogue.

\begin{figure*}
 \centering
 \includegraphics[width=0.75\textwidth,keepaspectratio]{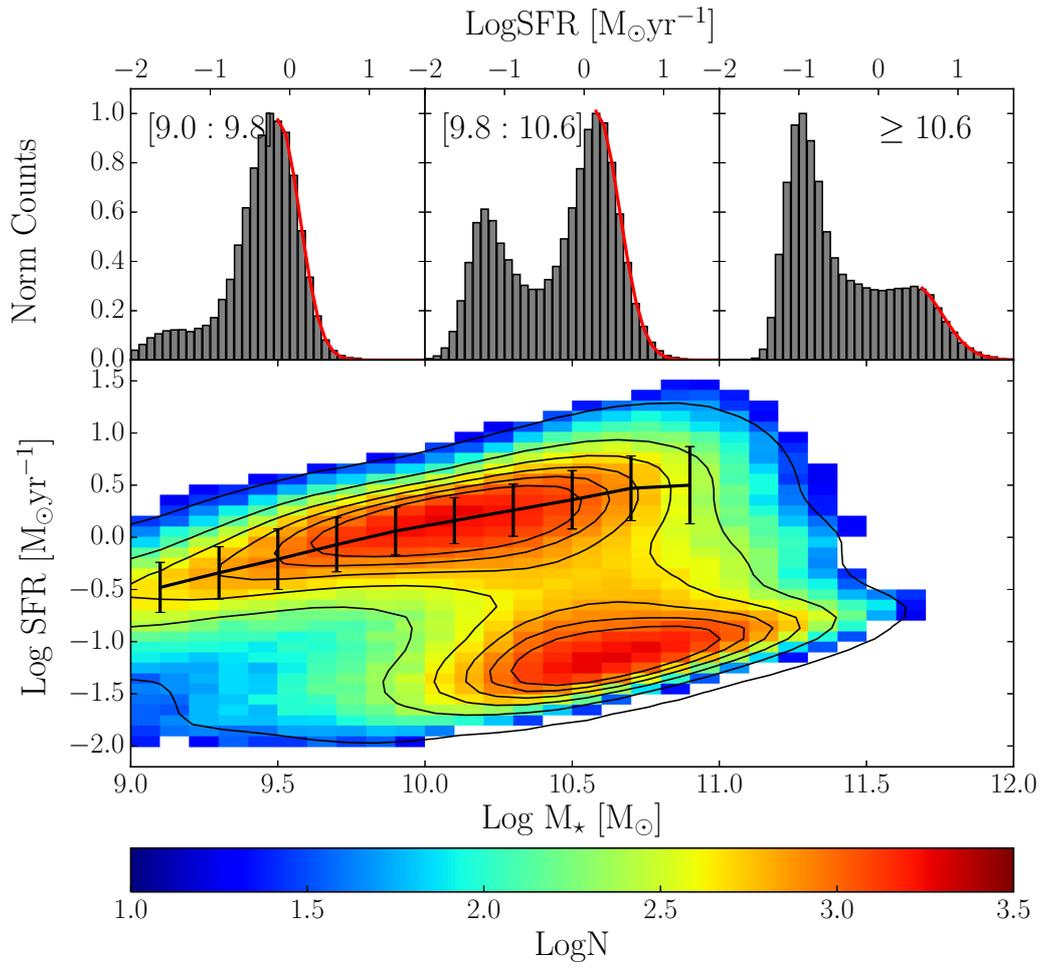}
\caption{Distribution of galaxies in $S_{ALL}$ sample in the LogSFR-LogM$_{\star}$ plane. {\it Top panels}: SFR distribution of galaxies in three different stellar mass bins. The red solid line is the result of the gaussian fit to the right-side of the SF peak. {\it Bottom panel}: LogSFR-LogM$_{\star}$ plane color-coded as a function of the number of galaxies in each bin. The contours encircle bins that host a number of galaxies from 50 to 1250, with step of 250. The MS of SFGs is indicated by the thick solid line, while the errorbars mark the 1$\sigma$ scatter.}
\label{fig1}
 \end{figure*}

\begin{figure*}
 \centering
 \includegraphics[width=0.85\textwidth,keepaspectratio]{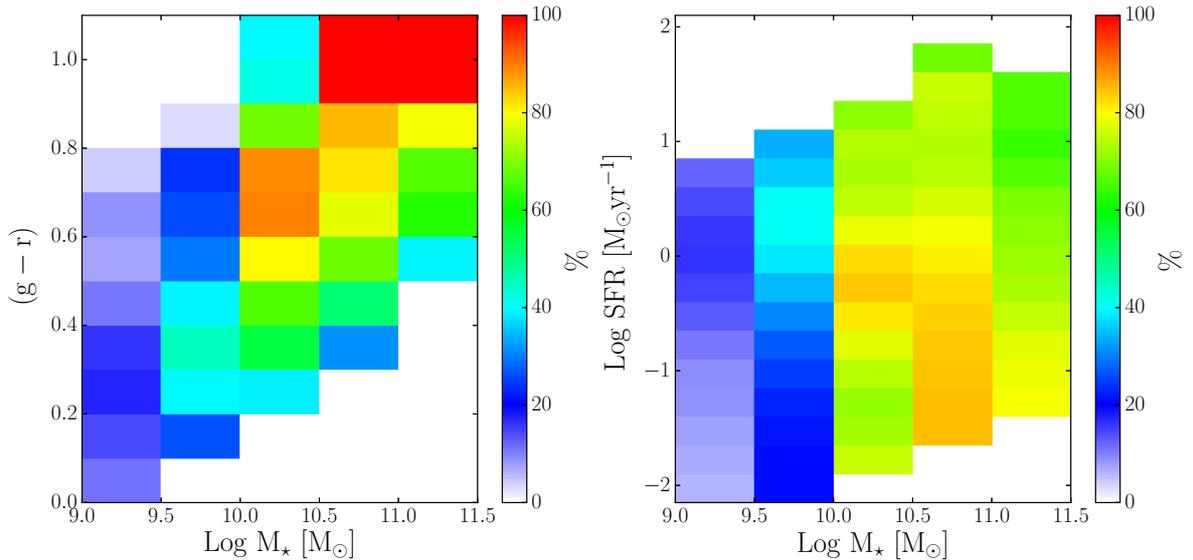}
\caption{ Completeness in the ($g-r$)-LogM$_{\star}$ plane (left panel) and LogSFR-LogM$_{\star}$ plane (right panel) of the spectroscopic $S_{ALL}$ sample with respect to the parent photometric $S_{phot}$ sample.}
\label{fig2}
 \end{figure*}

\subsection{The bulge/disc decomposition catalogue}
Galaxy structural parameters are taken from the bulge/disc decomposition of the \cite{2011ApJS..196...11S} catalogue (S11 hereafter). S11 apply the GIM2D code \citep{2002ApJS..142....1S} to the $g$ and $r$ filter images of 1,123,718 SDSS DR7 galaxies to obtain two-dimensional, point-spread function-convolved, bulge-disc decomposition. The simultaneous fit of the $g$ and $r$ images makes the decomposition robust against spurious effects. 
S11 provide structural parameters for three different fitting models: 1) a pure S\'ersic, 2) an exponential disc + De Vaucouleurs bulge, and 3) an exponential disc + free S\'ersic bulge model. In this work, we use the catalogue obtained from the exponential disc + De Vaucouleurs bulge model as the F-statistics reveal that SDSS images are not good enough to efficiently compute the S\'ersic index of the bulge and its structural parameters (see S11 for details). 
We use the B/T ratio computed from the $r$ band image and the magnitudes of the bulge and disc components, that are given as rest-frame, extinction and K corrected. \footnote{Extinction values are taken from the SDSS pipeline, while $K$ correction has been computed using version $4.2$ of \rm{k-correct} \citet{2007AJ....133..734B}}. To select robust subsamples of genuine bulge+disc systems we use the criterion $P_{PS}\le$0.32 (as suggested by S11), where $P_{PS}$ is the probability that a two component model is not statistically needed to fit the galaxy image.

\subsection{The galaxy sample}

We construct the master catalogue from the MPA-JHU SFR and M$_{\star}$ catalogue, after removing duplicates and objects with no SFR or M$_{\star}$  estimates, and cross-correlate it with the S11 catalogue. Less than 2\% of galaxies do not have a SFR or a stellar mass estimate due to low quality photometric or spectroscopic data. In the cross-correlation between the master and the S11 catalogues, only 8$\%$ of galaxies are lost due to minor differences in the selection criteria applied in S11. To limit our analysis to a local volume galaxy sample with relatively high spectroscopic completeness, we select galaxies that satisfy the following criteria:\\
\begin{itemize}
\item $0.02<z<0.1$;
\item LogM$_{\star}\ge$ 9.0 M$_{\odot}$;
\item no AGN as identified in the MPA-JHU catalogue;
\item good estimate of $z$ and SFR.
\end{itemize}

Our selection criteria lead to a final galaxy sample of $\sim$265,000 galaxies, $S_{ALL}$. In the bottom panel of Fig.~\ref{fig1} we show how galaxies in our $S_{ALL}$ sample populate the LogSFR-LogM$_{\star}$ plane. In the upper panels we show the distribution of the SFR of galaxies in three bins of stellar mass, to better visualise the typical bimodal distribution, and its dependency on stellar mass. The thick solid line and errorbars in the bottom panel of Fig. \ref{fig1} mark the position of the MS of SFGs and its $1\sigma$ scatter, respectively. Following the example of \cite{2015ApJ...801L..29R}, the MS and its scatter are computed as the mode and the dispersion of the SFR distribution in several stellar mass bins. We divide the sample in stellar mass bins of 0.2 dex in the range 9.0$\le$Log(M$_{*}$/M$_{\odot}$)$\le$11.0. In each bin, we fit with a gaussian to the right side of SFR distribution (see, as example, the red fits in the upper panels of Fig. 1), to avoid shift in the peak value due to the green valley population. We did not compute the MS values for masses M$_{\star}>10^{11}$M$_{\odot}$ since in this range the distribution is significantly non--gaussian \citep{2016MNRAS.455.2839E,2012ApJ...754L..29W} and can not be easily disentangled from the tail of the quiescent distribution. The mean ($\mu$) and sigma ($\sigma$) values of the gaussian fit are used as MS and scatter in the mid-point of each stellar mass bin, and are summarised in Table 1.

\begin{table}
\caption{ Main Sequence SFRs (first column) and dispersion values (second column) in bins of stellar mass, in Log(M/M$_{\odot}$) (third column).}
\label{tab1}
\centering
\begin{tabular}{ccc}
\hline
LogSFR$_{MS}$ [M$_{\odot} yr^{-1}$]  &  $\sigma_{MS}$ [M$_{\odot} yr^{-1}$] &  Log M$_{\star}$  [M$_{\odot}$]    \\
 \hline
-0.48    & 0.24 & 9.0:9.2\\    
-0.34    & 0.25& 9.2:9.4 \\
-0.21    & 0.29& 9.4:9.6\\
-0.07    &0.26& 9.6:9.8\\
0.06        & 0.23& 9.8:10.0\\
0.16      &0.22& 10.0:10.2\\
0.26        &0.25& 10.2:10.4\\
0.36      &0.28& 10.4:10.6\\
0.47      &0.31& 10.6:10.8\\
0.50  &   0.37& 10.8:11.0\\ 
 \hline
\end{tabular}
\end{table}

To check for biases introduced by our selection criteria, we study the completeness of our spectroscopic $S_{ALL}$ sample in the LogSFR-LogM$_{\star}$ plane, by comparing it with the parent photometric sample  drawn from the SDSS DR7 in the same region, redshift and stellar mass range ($S_{phot}$). The stellar masses for $S_{phot}$ galaxies without spectroscopic data are derived from the $z$ band absolute magnitude, M$_z$, using the best-fit LogM$_{\star}$-M$_z$ relation obtained for $S_{ALL}$ (Log(M$_{\star}$/M$_{\odot}$)=(-0.42$\pm0.1)M_z$+1.3($\pm$0.2)). On the other hand, obtaining a measure of the SFR without spectroscopic data is not trivial. We have verified that the five SDSS broad bands are insufficient to retrieve a reliable measure of the SFR through the SED fitting technique. Also, the addition of the GALEX UV bands does not help in obtaining a reliable dust correction. Indeed, the scatter between the SFR derived in this manner and the SFR derived from the dust-corrected $H{\alpha}$ emission of the MPA-JHU catalog exceeds 0.5 dex. To overcome this issue we use the rest-frame $(g-r)$ galaxy colour as a proxy for star formation activity.

\begin{figure}
 \centering
 \includegraphics[scale=0.46]{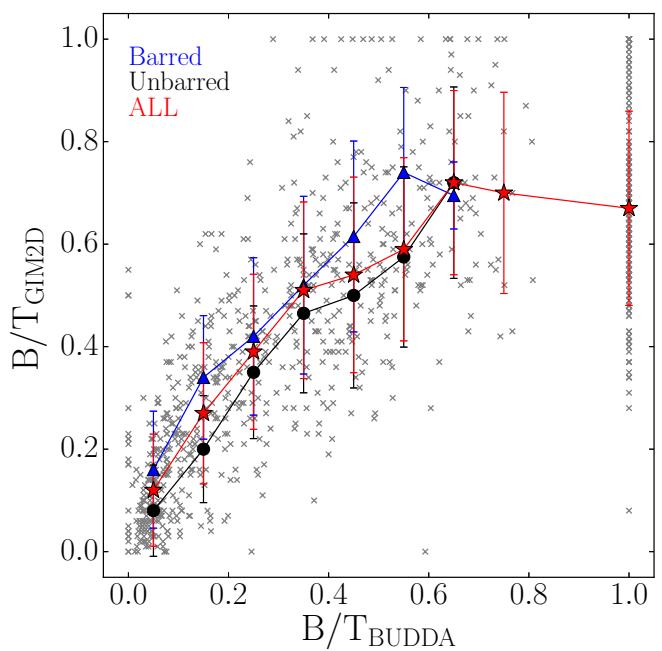}
\caption{Comparison between B/T$_{BUDDA}$ and B/T$_{GIM2D}$. The red stars are the median value of the B/T$_{GIM2D}$ in bins of 0.1 B/T$_{BUDDA}$ for all the galaxies that the two catalogues have in common, while the blue triangles and black circles are used for the subsamples of barred and unbarred galaxies, respectively. The error bars are given by the standard deviation in each bin.}  
\label{fig3}
 \end{figure}

The left panel of Fig. \ref{fig2} shows the $(g-r)$ - LogM$_{\star}$ plane color-coded as a function of the completeness level of $S_{ALL}$ with respect to $S_{phot}$. The completeness in each bin is estimated as the percentage of $S_{phot}$ galaxies that are also in the $S_{ALL}$ sample in the same bin. Completeness is below 50$\%$ for $10^{9.0}$M$_{\odot}<$M$_{\star}<10^{10.0}$M$_{\odot}$. In this stellar mass range it is larger for blue, star-forming galaxies than for passive, redder ones, due to the presence of strong emission lines. For M$_{\star}>10^{10.0}$M$_{\odot}$, the completeness is larger for redder galaxies, most likely due to the strong stellar continuum. To obtain the completeness in the LogSFR - Log$M_{\star}$ plane we proceed as follows. For every galaxy we get the completeness value by its $(g-r)$ colour and stellar mass (left panel). We then compute the completeness in a given bin of the LogSFR - Log$M_{\star}$ as the average completeness over all galaxies in that bin. The result is shown in the right panel of Fig. \ref{fig2}. Completeness varies between $5\%$ and $\sim 20\%$ for galaxies with $M_{\star}<10^{9.5}M_{\odot}$, where it is larger for blue, star-forming galaxies due to the presence of strong emission lines. For stellar masses in $10^{9.5-10.0}M_{\odot}$, the completeness of the spectroscopic catalogue reaches $40\%$ on the MS and in its upper envelop, while it decreases to $20\%$ for passive galaxies. For galaxies with $M_{\star}>10^{10}M_{\odot}$, completeness is always higher than $60\%$ and it reaches values of $\sim80\%$ on the MS, and is even larger in the passive region with $M_{\star}\sim10^{10.5-11.0}M_{\odot}$ where galaxies are characterised by a strong continuum.

  \begin{figure}
 \centering
 \includegraphics[scale=0.5]{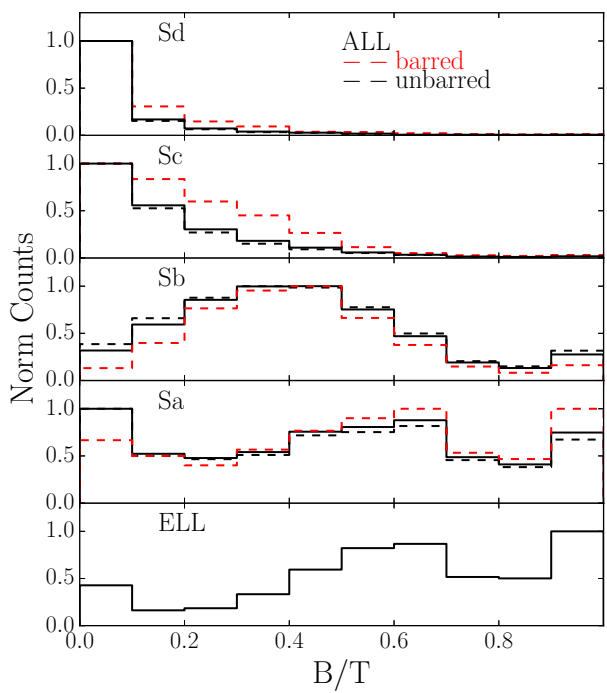}
\caption{ Distribution of the S11 B/T in the different morphological classes defined by Galaxy Zoo 2 (from top top to bottom: Sd, Sc, Sb, Sa, and ellipticals). The histograms have been normalised to the peak. The red dashed histograms describe barred spirals, while the black dashed ones unbarred spirals. The whole population of barred and unbarred galaxies is described by the black solid histograms.}
\label{fig4}
 \end{figure}

 \subsubsection{Bulge/disc decomposition reliability: comparison with BUDDA and  Galaxy Zoo}
\label{sec_s11_budda}
To check the reliability of the S11 B/T classification, we compare it with different morphological classifications, in particular: BUDDA \citep[Bulge/disc Decomposition Analysis]{2004ApJS..153..411D}; and {\it Galaxy Zoo} \citep{2008MNRAS.389.1179L}. Both classifications take into account the presence of bars. Thus, this comparison  is particularly interesting in order to understand how the presence of a bar can bias the S11 B/T estimate.

BUDDA is a {\sc{fortran}} code that performs multicomponent decomposition of galaxies. Bulges are fitted with S\'ersic profiles, and discs with exponentials. Unlike in GIM2D, in addition to bulge and disc, bars are also modelled with a S\'ersic profile. We make use of the public catalogue of structural parameters presented in \citet[][G09 hereafter]{2009MNRAS.393.1531G}, where the BUDDA code has been applied to a sample of 946 spectroscopically confirmed, nearly face on galaxies, in the range $0.02\le z\le 0.07$ and M$_{\star}>10^{10}$M$_{\odot}$. The final sample of G09 is clean and optimal for bulge/disc decomposition as the quality of each image has been inspected by eye and the redshift range is low. The S11 and G09 catalogues have 910 galaxies in common (correlation radius $r$ = 3''). We compare the B/T ratios of S11 (B/T$_{GIM2D}$) with the G09 values (B/T$_{BUDDA}$). Both B/T values have been computed from the $r-$band images.  Results are shown in Fig. \ref{fig3}.  The red stars are the median values of the whole sample in bins of 0.1 in B/T, and the errors are computed as  the standard deviation in each bin. We find an overall agreement between the two methods at least below B/T$\sim$0.6. GIM2D tends to find slightly larger B/Ts with respect to BUDDA. 
The blue triangles mark the subsample of galaxies that are classified as "barred" in BUDDA, while the black circles indicate the "unbarred" subsample. A very good agreement in B/T$_{BUDDA}$ and B/T$_{GIM2D}$ is found for unbarred galaxies, while for barred galaxies, GIM2D tends to estimate larger values of the B/T - almost double at B/T$< 0.1$ and $20-30\%$ larger at larger B/Ts. Despite this, the overall effect is not dramatic as B/T$_{GIM2D}$ for the whole sample (red stars) does not increase more than 25\% due to the inclusion of barred galaxies. This is because of their relatively low fraction $\sim$ 30\%. Above the B/T$\sim$0.6 threshold, B/T$_{GIM2D}$ tends to flatten to an average value of $\sim$0.7. This is consistent with the tendency of GIM2D to find an overabundance of small discs in spheroidal galaxies, as pointed out in S11.

We compare GIM2D with the second release of Galaxy Zoo \citep{2008MNRAS.389.1179L}, a project aimed at classifying "bye-eye" the morphology of galaxies in the SDSS Main Galaxy Sample \citep{2002AJ....124.1810S}. The catalog provides morphological classification only for $\sim50\%$ of the parent photometric sample drawn from the SDSS. The remaining half of the galaxy sample is classified with ''uncertain'' morphology. In the second release \citep[GZ2;][]{2013AAS...22134005W} ellipticals are classified as {\it round}, {\it cigar-like}, or {\it in-between}.  Spirals are classified in four classes as a function of the bulge prominency: {\it Sa} if they have dominant bulge, {\it Sb} for obvious bulges, {\it Sc} if the bulge is just noticeable, and {\it Sd} if there is no bulge. In addition, spirals are also divided in {\it barred} and {\it not-barred}. Since edge-on spirals have separate morphological classes in GZ2, we restrict this analysis to galaxies that, in S11, have a disc inclination angle smaller than 70$^{\circ}$ (where $i$ = 90$^{\circ}$ for an edge-on disc). In Fig. \ref{fig4}, the B/T distribution for each sub-category is shown.  GZ2 ellipticals, considered as a whole regardless of the shape, are characterised by a B/T distribution peaking at B/T$\sim 1$ and with a second peak at B/T$\sim0.6$. We also observe a third much less significant peak at B/T of 0. Approximately 60\% of GZ2 ellipticals with B/T$<0.8$ have $P_{PS}\le 0.32$, thus making the disc component  statically significant.  These galaxies could be S0 galaxies that are visually misclassified as ellipticals \citep{2013AAS...22134005W,2011A&A...525A.157H}. Not surprisingly, for GZ2 {\it Sa} galaxies, i.e. consistent with an intermediate morphology template (bulge dominated spiral), the B/T distribution is wide with 3 peaks, at B/T of 0, $\sim$0.5 and 1, respectively. 
However, we point out that {\it Sa}  galaxies account only for less than $\sim 1\%$ of the GZ2 sample. For {\it Sb}, {\it Sc}, and {\it Sd} the B/T distributions reveal a very good agreement between GZ2 and GIM2D classification. Also, it is clear from Fig.\ref{fig4} that the presence of a bar (red dashed line histograms) results in an overestimate of the GIM2D B/T, in particular for {\it Sc} and {\it Sd} galaxies. 

We point out that neither G09 or GZ2 catalogues are suitable for our analysis. While BUDDA does not provide the required statistics, in GZ2, the visual classification is available only for half of the parent galaxy subsample drawn from the SDSS, leading to a high level of incompleteness. Only S11simultaneously provides the required statistics and completeness.

 \begin{figure*}
 \centering
  \includegraphics[width=0.9\textwidth, keepaspectratio]{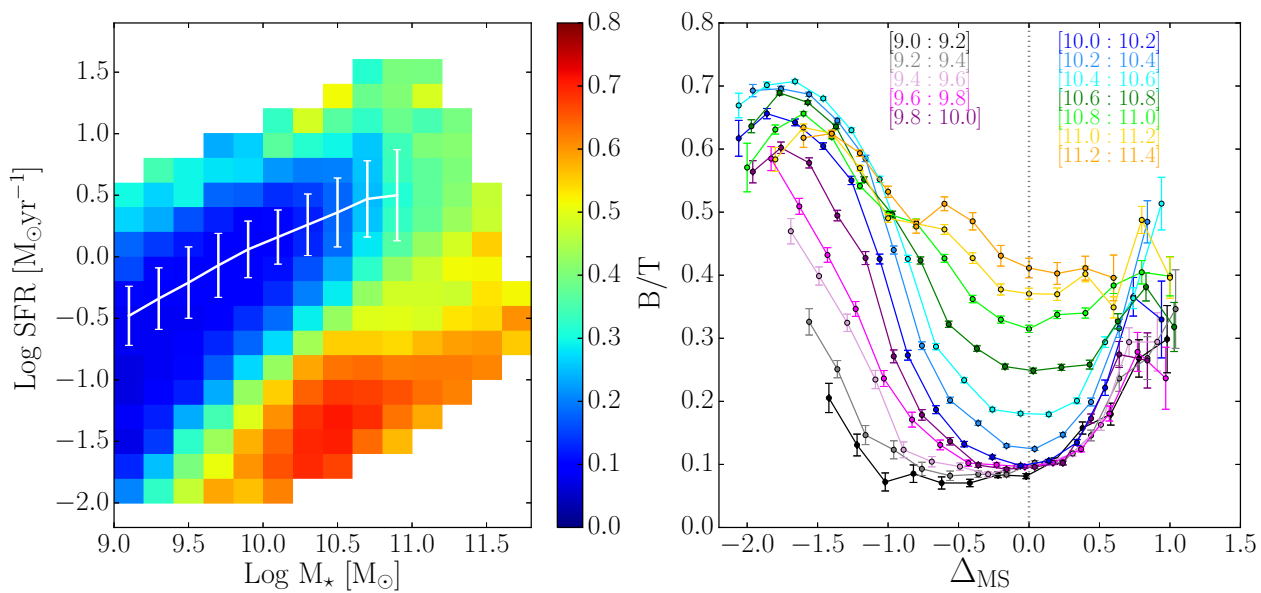}
\caption{$Left\ panel$: LogSFR-LogM$_{\star}$ plane color coded according to the weighted average B/T in the bin. The white line represent the location of the MS of star forming galaxies, and the errorbars its dispersion. $Right\ panel$: B/T ratio as a function of $\Delta_{MS}$ (distance from the MS) in different stellar mass bins. The dotted vertical line mark the position of the MS. Errorbars are obtained via bootstrapping.}
\label{fig9}
 \end{figure*}

The drawback of using S11 is the uncertainty of bulge/disc decomposition at relatively large values of B/T. Indeed, the comparison with BUDDA and GZ2 shows that S11 find an overabundance of double component systems among galaxies that are classified as pure spheroidal by BUDDA and by the visual classification. \cite{2013ApJ...779..162C} suggest that, among galaxies with B/T$>$0.5, only the ones with $P_{PS}<0.32$ have a reliable the bulge/disc decomposition. However, such a drastic cut would imply selecting only half of the S11 sample at B/T$>$0.5, leading to significant selection effects in our analysis. To overcome this problem, we adopt the following approach. {When measuring the mean B/T in the LogSFR-LogM$_{\star}$ plane, we use a weighted mean, where the weights are based on the errors provided by the S11 catalog for B/T. Including this error in the weighted mean allows us to take into account the uncertainty due to low S/N and resolution issues. This approach also allows us to propagate the uncertainty of the morphological classification throughout our analysis. As an alternative approach, and to verify the robustness of our results, we create a reference sample ($S_{ref}$) where we keep the B/T as given by S11 for all galaxies with $P_{PS}\le0.32$. Instead, where $P_{PS}>0.32$, for values of B/T$<$0.5 and B/T$>$0.5, we consider the galaxy either a pure disc ($B/T=0$) or pure spheroidal ($B/T=1$), respectively. The results based on $S_{ref}$ are shown for comparison in Appendix \ref{Sref}. When analysing the colours of the individual components in the LogSFR-LogM$_{\star}$ plane, we use a sample with robust double component classification by imposing the cut at $P_{PS}<0.32$.}

\section{Results}
In the following sections, we study the galaxy B/T and bulge/disc colours in the LogSFR-LogM$_{\star}$ plane. The purpose of this analysis is to better understand the drivers of the MS scatter, and the main differences in galaxy properties between SF and passive galaxies. 

   \begin{figure}
 \centering
 \includegraphics[width=0.48\textwidth, keepaspectratio]{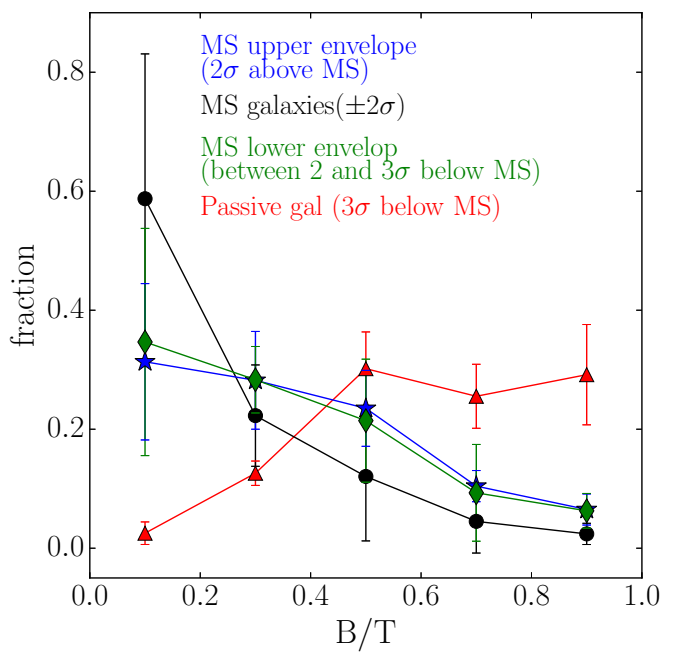}
\caption{Fraction of galaxies as a function of the B/T within the MS (black circles), in the upper envelope of the MS (blue stars), and in the green valley and in the passive region (red triangles). The fraction is estimated as the mean in several bin of stellar mass from $10^{10}$ to $10^{11}$ M$_{\odot}$. The error bars show the dispersion around the mean. The MS  upper envelope and the lower envelope (green valley) share the same distribution of galaxies as a function of the B/T: $\sim$ 40\% of the population has B/T $>$ 0.4 versus the 15\% of the galaxies in the MS.}
\label{evidence}
 \end{figure}

 \subsection{The B/T across the LogSFR-LogM$_{\star}$ plane}
 \label{sec_bt}
 
In Fig. \ref{fig9} we show the distribution of the B/T values of $S_{ALL}$ in the LogSFR-LogM$_{\star}$ plane (left panel). Each bin is color coded according to the weighted average of the B/Ts in that bin. The width of each bin is 0.2 in both Log$M_{\star}$ and Log SFR, in order to account for the average errors with a minimum of 20 galaxies per bin. The black solid line is the MS and the error bars show the dispersion around it (Table 1). 

   \begin{figure*}
 \centering
 \includegraphics[width=0.47\textwidth, keepaspectratio]{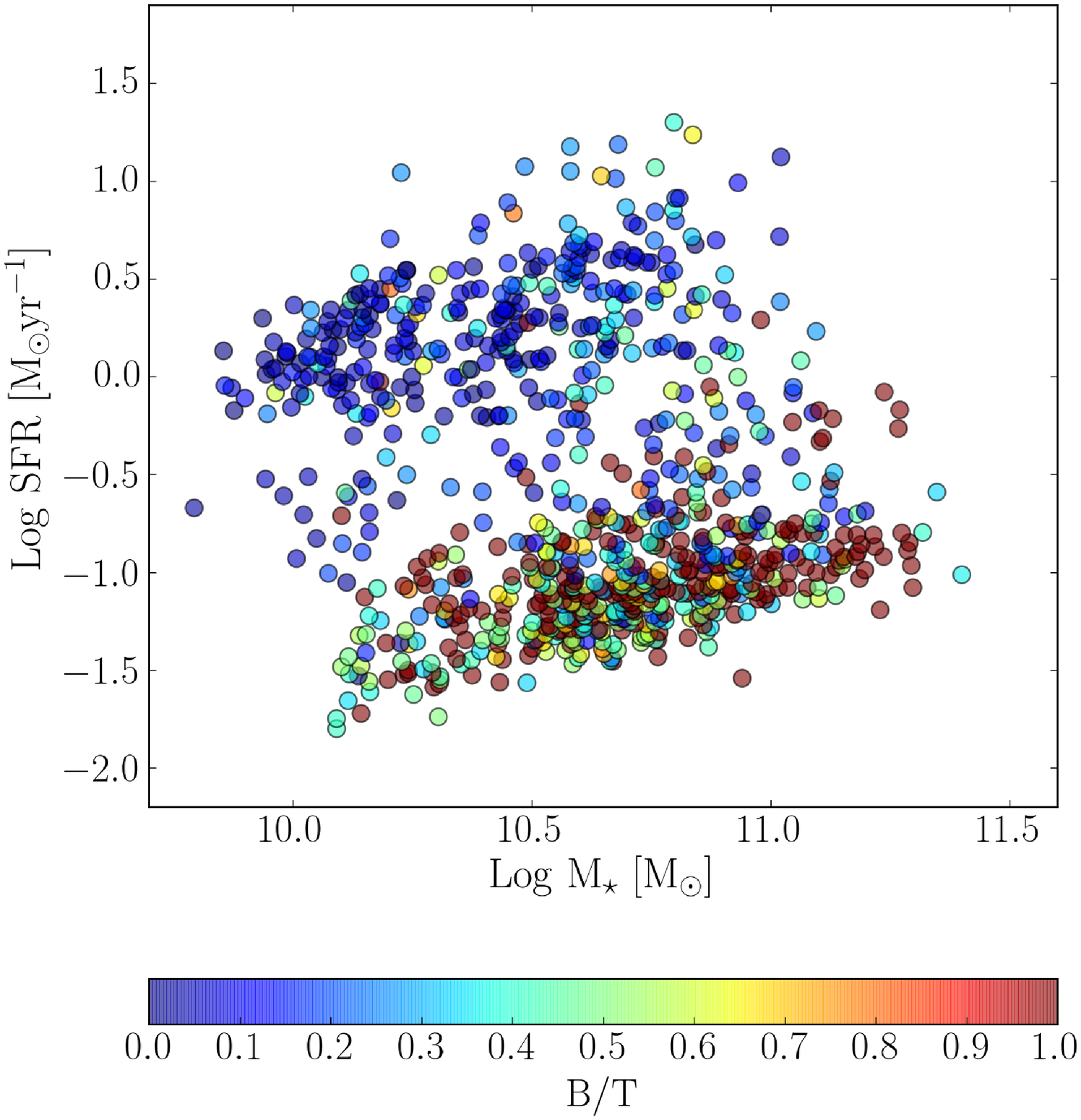}
 \includegraphics[width=0.48\textwidth, keepaspectratio]{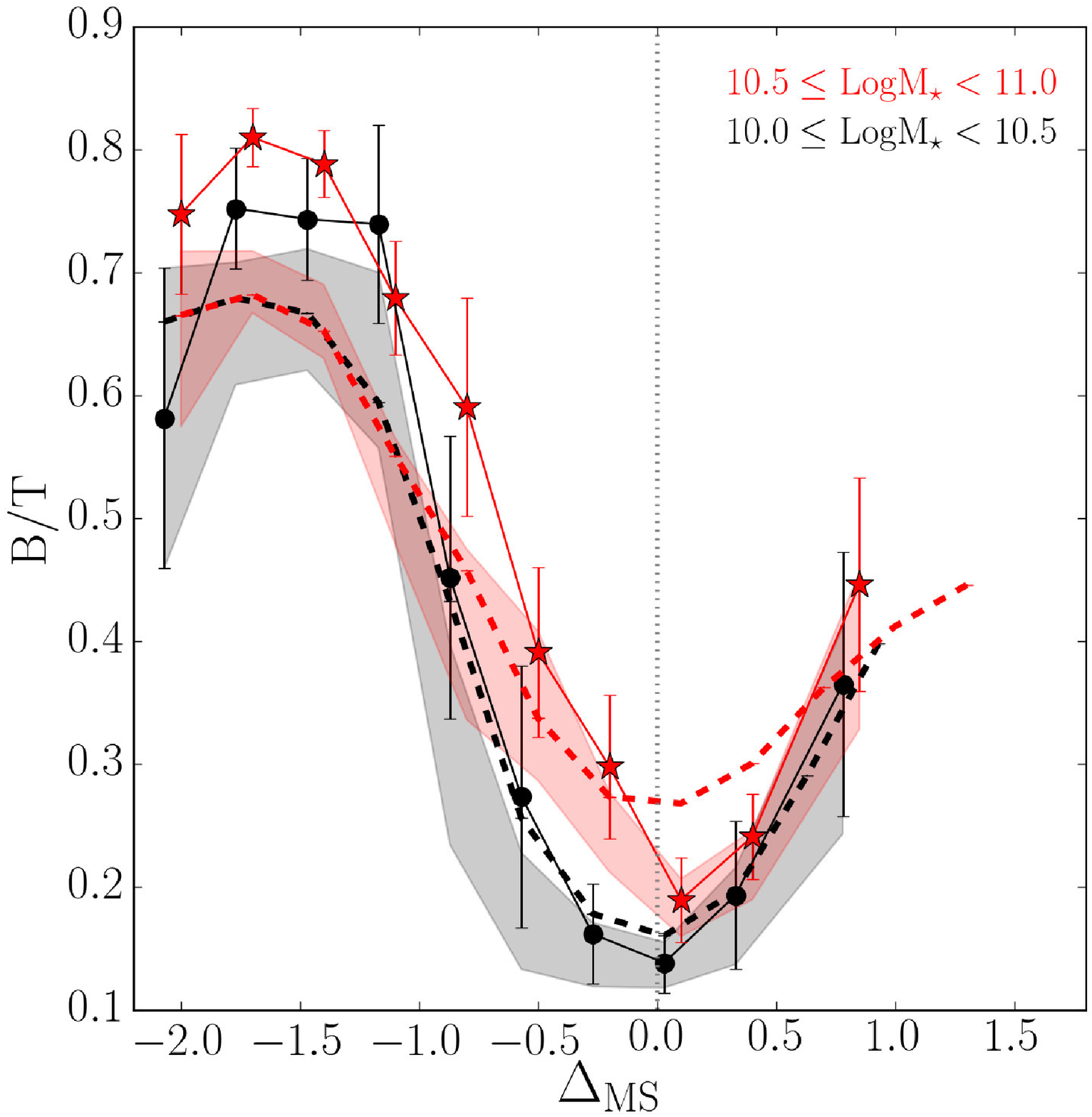}
\caption{ {\it Left panel}: LogSFR-LogM$_{\star}$ plane for galaxies in the bulge-disc decomposition catalogue of G09. Each marker is color-coded accordingly to the B/T value of the galaxy. {\it Right panel}: B/T as a function of the distance from the MS, $\Delta_{MS}$, for the G09 whole sample (shaded region), and for the G09 subsample of unbarred galaxies (solid lines). The sample has been divided in 2 bins of stellar mass:  $10.0\le$ Log(M$_{\star}$/M$_{\odot}$)$<$10.5 (in black), and $10.5\le$ Log(M$_{\star}$/M$_{\odot}$)$<$11.0 (in red). The B/T distributions computed for the $S_{ALL}$ sample are also shown for comparison in dashed lines. The solid vertical lines marks the location of the MS. Errorbars are obtained via bootstrapping.}
\label{fig.g09}
 \end{figure*}

Fig. \ref{fig9} confirms the result already obtained by Wuyts et al. (2011): the MS is populated mainly by disc dominated galaxies and the quiescence region by bulge dominated systems. Intermediate B/T values are found in the green valley and in the upper envelope of the MS. We also confirms previous results that the mean B/T on the MS increases as a function of the stellar mass, going from 0.16 at 10.2$<$LogM$_{\star}<$10.4, to 0.21 at 10.4$<$LogM$_{\star}<$10.6, 0.26 at 10.6$<$LogM$_{\star}<$10.8, and 0.32 at 10.8$<$LogM$_{\star}<$11.0.  This implies that the MS above LogM$_{\star}=$10.2 begins to be populated by double component galaxies, consistent with previous findings (Lang et al. 2014, Bluck et al. 2014, Erfanianfar et al. 2016). When approaching the quiescence region at $\gtrsim$1 dex below the MS, the B/T further increases to reach an average value between $0.6-0.7$, for galaxies with LogM$_{\star}>$9.6. For less massive galaxies, the average B/T in the passive region is in the range 0.3-0.5. However, this is the region most affected by completeness issues as shown in Fig. \ref{fig2}. Thus, it is very likely that a strong selection effect in favour of emission line galaxies is biasing the mean value of the B/T in the weighted mean.

\begin{figure*}
 \centering
 \includegraphics[width=0.9\textwidth,keepaspectratio]{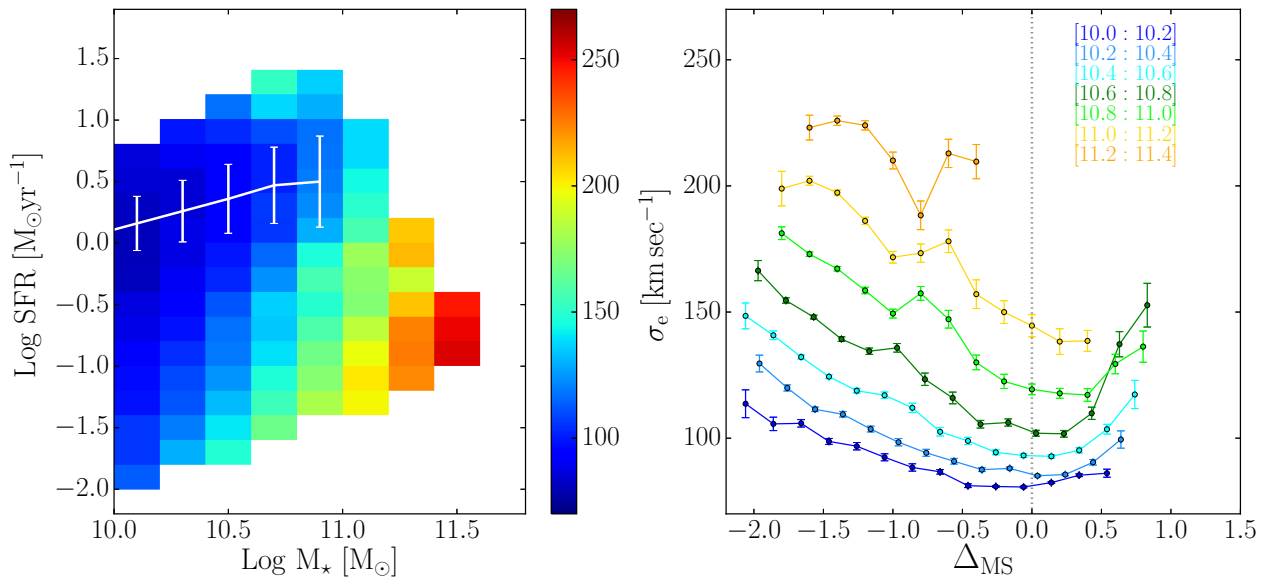}
\caption{$Left\ panel$: distribution of galaxies in the LogSFR-LogM$_{\star}$ plane, in which each bin (0.2M$_{\odot}$ in M$_{\star}$ and 0.2 M$_{\odot}yr^{-1}$ in SFR) is color coded according to the weighted average value of the velocity dispersion $\sigma_{e}$ in the bin. The white line represents the location of the MS of star forming galaxies. $Right\ panel$: $\sigma_{e}$ as a function of the distance from the MS, $\Delta_{MS}$, in 6 different bins of stellar mass, each 0.2M$_{\odot}$ wide. The different bins of stellar mass are indicated by different colours, as defined in the legend of the image, while the grey dotted vertical line represents the location of the MS. The errorbars are obtained via bootstrapping. }
\label{fig.vdisp}
 \end{figure*}

A deeper insight is obtained from a careful analysis of the relation between the residuals around the MS, $\Delta_{MS}$=LogSFR$_{MS}$-LogSFR$_{gal}$, (where SFR$_{MS}$ is the SFR on the MS and SFR$_{gal}$ is the galaxy SFR), and the mean B/T in different stellar mass bins as shown in the right panel of Fig. \ref{fig9}. Contrary to Wuyts et al. (2011), where they find a roughly constant value of the S\'ersic index $n \sim 1$ around the MS, we observe that the B/T-$\Delta_{MS}$ relation exhibits a parabola-like shape within 3$\sigma$ from the MS, with the MS coinciding with the lowest B/Ts at any stellar mass. 
Differently from previous results (Wuyts et al. 2011, Cheung et al. 2012), we find that star forming blue galaxies with a bulge dominated morphology are not outliers in the diagram but are just populating the upper envelope tail of the MS.
To further investigate how galaxies with different morphologies populate the plane, we study the distribution in the entire stellar mass range of the B/T for galaxies above the MS ($> 2 \sigma$), within the MS ($\pm2\sigma$), in the lower envelop of the MS ($ 2 -3 \sigma$ below), and in the passive region (3$\sigma$ below). As shown in fig. \ref{evidence}, the MS (black line) is dominated by pure disk galaxies where $\sim$60\% of the population has B/T$<$0.2. The large error associated to MS galaxies with B/T$\le$0.2 reflects the increase of the B/T along the MS. In the upper envelope of the MS (blue stars), disk galaxies make up only $\sim$30\% of the entire galaxy sample, favouring a population of intermediate B/Ts. Similarly, in the lower envelop of the MS $\sim35\%$ of galaxies are pure disks. Galaxies in the lower and upper envelope of the MS share the same distribution of galaxy B/T . Among passive galaxies (red triangles), $\approx 70\%$ have B/T$\le$0.8. As discussed earlier, this fraction is contaminated by true spheroidal galaxies for which GIM2D finds a spurious disk. Nevertheless, there is a significant population of passive galaxies with a secure bulge+disk morphology where $\approx45\%$ of all passive galaxies have B/T$\le$0.8 and $P_{PS}<0.32$. We also notice that the fraction of passive galaxies with B/T$<0.2$ is negligible, indicating that red pure disks, if at all present, are outliers of the galaxy population.

These trends of B/T in the LogSFR-LogM$_{\star}$ are also confirmed by using the $S_{ref}$ sample (see Appendix A). As an additional test, we repeat the analysis by limiting the sample to galaxies with high S/N photometric data and secure S11 classification, by selecting galaxies with a disk scale length larger than 2 arcsec to avoid resolution issues (as in Cheung et al. 2013). All the tests confirm the results presented in Fig. \ref{fig9}.

\subsection{The effect of bars}

In Sec.\ref{sec_s11_budda} we conclude that S11 and G09 show good agreement for B/T$<0.6$, and that the presence of a bar leads to an overestimation of the S11 B/T of  $10-25\%$. However, the gradient observed from the peak of the MS towards the upper and lower envelope of the relation leads to an increase of the mean B/T by a factor $\sim 3-5$, depending on the stellar mass bin. Thus, the  increase of the mean B/T across the MS is much larger than the overestimation of the B/T due to bars. Nevertheless, bars could be preferentially located in certain regions of the LogSFR-LogM$_{\star}$ plane, thus affecting the average B/T value of that region.  

To check this, we perform the same analysis as in Fig. \ref{fig9} on the G09 sample (see left panel of Fig. \ref{fig.g09}).  We observe a $\sim$ 25\% milder increase of the mean B/T along the MS with respect to the result of Fig \ref{fig9} (mean B/T value of 0.13, 0.19, and 0.23 at stellar masses of $10^{10-10.4}$, $10^{10.4-10.8}$, and $10^{10.8-11.2}$ M$_{\odot}$, respectively), consistent with the overall effect of overestimation of the S11 B/T due to the inclusion of barred galaxies.  Due to the lower statistics, the upper and the lower envelopes of the MS in G09 are poorly populated. Thus, we limit the analysis of the mean B/T as a function of the distance from the MS to two stellar mass bins, above and below the median mass of the sample ($10^{10.25} $M$_{\odot}$). We perform the  analysis for the whole G09 sample and for the ``unbarred'' G09 sample (deprived of barred galaxies). The right panel of Fig. \ref{fig.g09} shows the mean B/T as a function of $\Delta_{MS}$ for the two G09 samples in comparison with the mean relation obtained in the same stellar mass bins in S11. The error bars are estimated via bootstrapping technique. We find a perfect match of all relations in the MS region. This indicates that bars do not play a crucial role in the gradient of B/T across the MS. Only in the quiescence region, as expected, the S11 relation shows lower mean B/T with respect to G09 due to the underestimation of S11 B/T for spheroidal galaxies (Fig. \ref{fig3}). In addition the barred galaxy sample of G09 shows that the B/T-$\Delta_{MS}$ for barred galaxies exhibits the same parabola-like shape. Thus, barred galaxies follow the same trend. This suggests, that the presence of a bar, on average plays a second order effect with respect to the galaxy morphology.

\subsection{Velocity dispersion in the LogSFR-LogM$_{\star}$ plane}
\label{disp}

To investigate the similarities and the differences of the bulgy galaxy population in the upper and lower limit of the MS, we use the bulge velocity dispersion, $\sigma$, and the SDSS concentration parameter, $R_{90}/R_{50}$. $\sigma$ and $R_{90}/R_{50}$ are often used to discriminate between classical and pseudo bulges (see e.g. G09). Classical bulges resemble spheroidal galaxies for their colours and scaling relations. Pseudo-bulges are, instead,  outliers in the Kormendy relation \citep[][ G09]{1977ApJ...218..333K}. They tend to be blue and with low velocity dispersion with a S\'ersic index close to $\sim$2.

To retrieve the bulge velocity dispersion, we use the estimate of the stellar velocity dispersion provided by MPA-MJU catalog, measured by fitting the stellar absorption features of the SDSS galaxy spectra. The aperture velocity dispersion, $\sigma_{ap}$, is computed inside the SDSS fiber, characterised by a 3 arcsec diameter. Such an aperture samples different galaxy parts as a function of: redshift, galaxy size, and morphology. In order to sample only the bulge, we select galaxies with B/T$\ge$0.5, and galaxies with $P_{PS} \le 0.32$ if B/T$<$0.5. In order to exclude contamination by disc rotation, we limit our analysis to nearly face-on galaxies, with disc inclination $i < 30^{\circ}$.\footnote{the disk inclination angle $i$ is an output of the bulge/disk decomposition and it is 0$^{\circ}$ for face-on galaxies, $90^{\circ}$ for edge-on galaxies.} 

Given the spectral resolution of the SDSS spectrograph and the typical S/N values, velocity dispersion estimates smaller than $\sim$70 km/sec are not reliable. Thus, to all galaxies with $\sigma_{ap}<$ 70 km/sec, we assign a fixed upper limit of $\sigma_{ap}$ = 70 km/sec, with an error of $\pm$70 km/sec to be used for the weight calculation in the weighted mean. We also exclude galaxies with $\sigma_{ap}>420$ km/sec\footnote{The velocity dispersion measurements distributed with SDSS spectra use template spectra convolved to a maximum sigma of 420 km/s.} and with low S/N continuum ($<3$). We apply the aperture correction of Jorgensen et al. (1995) to $\sigma_{ap}$ in order to compute $\sigma_{e}$: the velocity dispersion within the effective radii of the galaxy, $r_e$ (that we take from SDSS database). In the redshift range considered here, the average aperture correction is $\sim14\%$. 

\begin{figure*}
 \centering
 \includegraphics[width=0.9\textwidth,keepaspectratio]{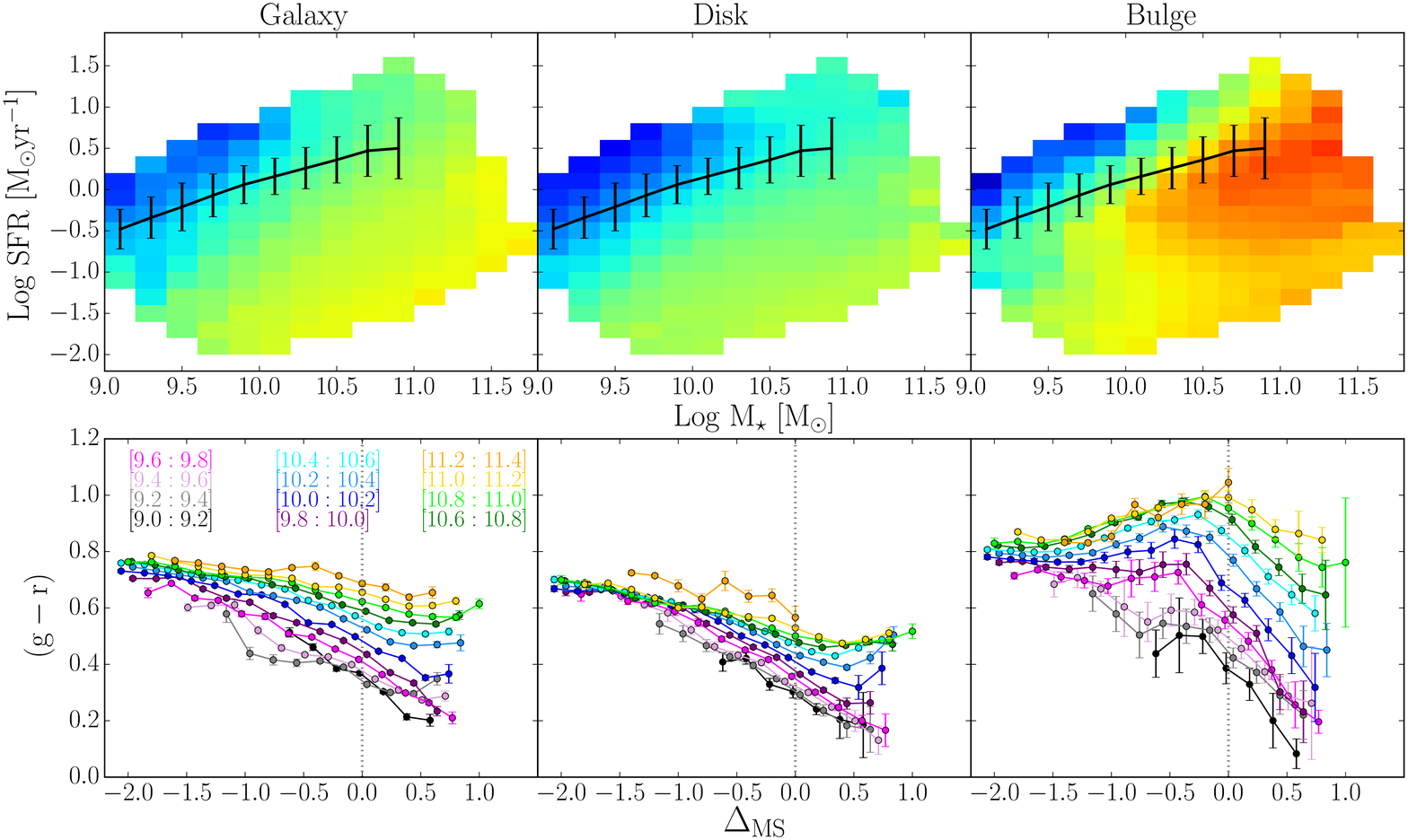}
 \includegraphics[trim={17cm 0 5 0cm},clip,scale=0.25]{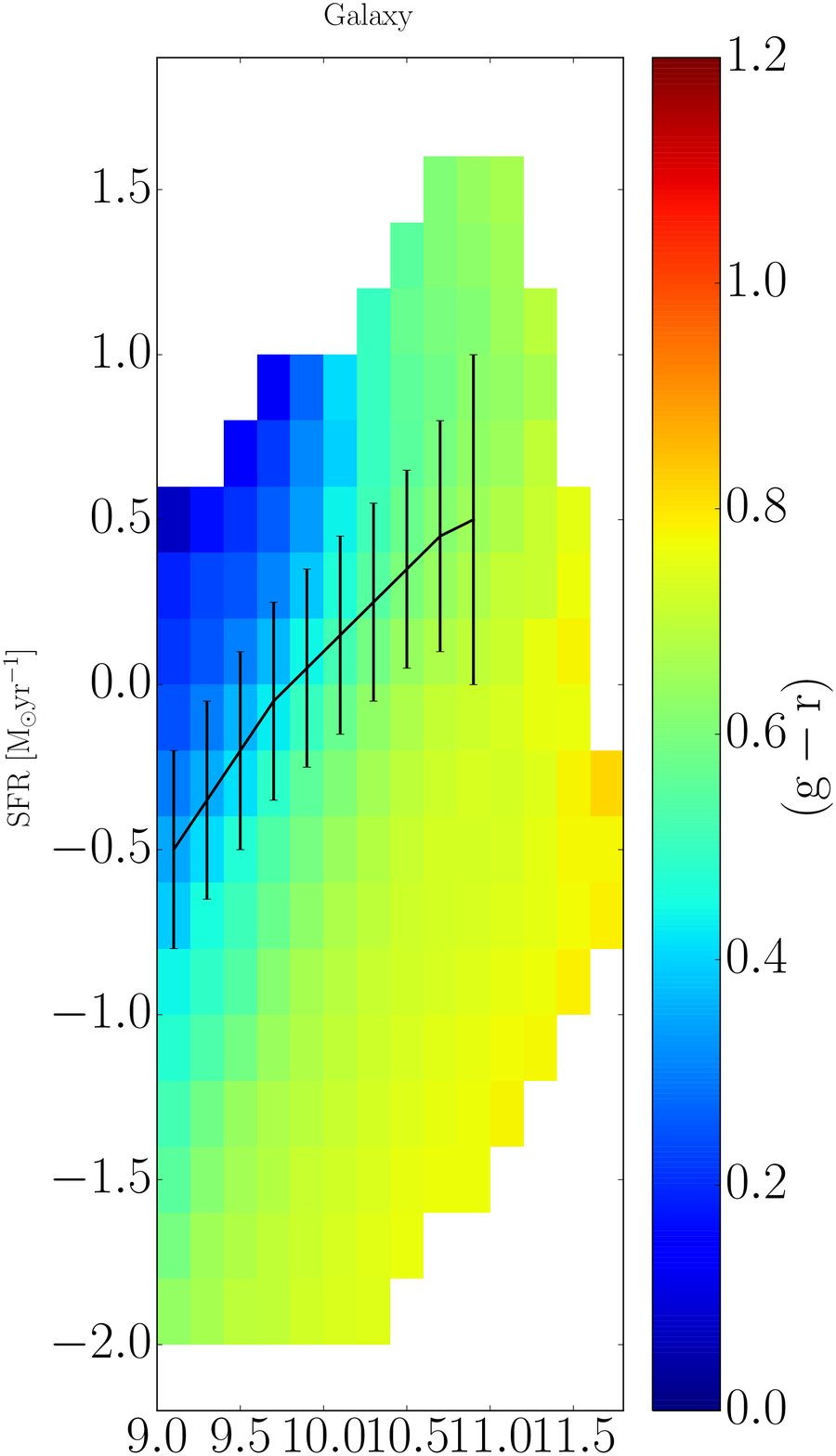}
\caption{Galaxy (left panels), bulge (central panels), and disc (right panels) colours in the LogSFR-LogM$_{\star}$ plane. In the upper panels, the distribution of galaxies in the LogSFR-LogM$_{\star}$ plane is color coded according to the weighted average $(g-r)$ colour in each bin. The bins are 0.2M$_{\odot}$ in Log$M_{\star}$ and 0.2M$_{\odot}yr^{-1}$ in LogSFR. In the bottom panels, the $(g-r)$ color is shown as a function of the distance from the MS, $\Delta_{MS}$, in different bins of stellar mass, represented by different colours (same as Fig. \ref{fig9}). The error bars in the lower panels are obtained via a bootstrapping. }
\label{fig14}
 \end{figure*}

Fig. \ref{fig.vdisp} shows the distribution of galaxies in the LogSFR-LogM$_{\star}$ plane color coded as a function of the average $\sigma_{e}$ in each bin. The error bars in Fig. \ref{fig.vdisp} are estimated via a bootstrapping analysis. The behaviour of $\sigma_{e}$ across and along the MS is qualitatively similar to the B/T ratio. At a fixed stellar mass, $\sigma_{e}$ has its minimum on the MS, increases towards upper and lower envelope of the MS, and reaches its maximum in the quiescence region. Also, the velocity dispersion of MS galaxies increases for increasing stellar mass along the MS, resembling the B/T behaviour. We point out that the exclusion of pure-disc galaxies, that populate the core of the MS, naturally leads to a flatter parabola-like shape for the $\sigma_{e}-\Delta_{MS}$ distribution as a function of the mass.

The same trend in Fig. \ref{fig.vdisp} is also observed for the concentration parameter, which is one of the most reliable discriminant (G09) between bulges and pseudo-bulges at given B/T (see Appendix \ref{append_comp}). In the lower and upper envelope of the MS the mean value of $R_{90}/R_{50} \approx 2.5$ is consistent with the mean for classical bulges with low values of D4000, as estimated in the same mass range by G09 (see Fig. 20 of G09). Low values of  the D4000 index would indicate a relatively young age of the stellar population. Values of $R_{90}/R_{50} \approx 2$, consistent with pseudo-bulges, are found only in the core of the MS. In the quiescence region, the mean $R_{90}/R_{50} \approx 3$ is consistent with classical bulges with high values of D4000, i.e older stellar populations.
Thus, we conclude that lower and upper envelopes of the MS shares the same B/T, bulge velocity dispersion and concentration parameter distributions. Both are populated by intermediate morphology galaxies with classical bulges. The similarity of the these distributions could suggest that the scatter around the MS is characterised by different evolutionary stages of the same galaxy population, while the MS itself is mainly populated by pure disc galaxies.

\subsection{Bulge and disc colours in the LogSFR-LogM$_{\star}$ plane}
\label{colours}

The SDSS spectroscopic dataset does not provide any spatial information. Thus to understand if there is a connection between the SF activity of the individual galaxy components (bulge and disc), and the star formation of the galaxy as a whole, we use the color as a proxy of the SFR. For this purpose we limit this analysis to the $S_{ALL}$ sample with secure double morphological component, applying the cut at $P_{PS} \le 0.32$.

In Fig. \ref{fig14} we show the $(g-r)$ colour of the galaxy (upper left panels), bulge (upper central panels), and disc (upper right panels) in the LogSFR-LogM$_{\star}$ plane. The upper panels are color coded as a function of the galaxy or galaxy component color. The color in each bin is obtained with a weighted mean. The bottom panels show the dependence of the mean colour of the whole galaxy (left panel), the disc (central panel) and the bulge (right panel) on the distance from the MS. 

The galaxy and the disc colours follow similar trends where at any stellar mass bin, they anti-correlate with the distance from the MS, getting progressively bluer from the quiescence region to the upper envelope of the MS. The anti-correlation is steeper for the whole galaxy color than for the disc component. In both cases, the relation flattens progressively towards highest stellar mass bin. The bulge colour shows a different behaviour instead. Up to $\sim 10^{10} $M$_{\odot}$, the bulge color also anti-correlates with the distance from the MS, getting bluer from the passive region to the upper envelope of the MS. In this stellar mass range bulges above the MS are as blue as their discs. However, above $\sim 10^{10} $M$_{\odot}$, the relation reverses with the bulge color getting redder from the passive region up to the MS. After reaching its reddest value, the bulge turns slightly bluer towards the upper envelope of the MS. However, the bulge color is always redder than the disc even in the upper envelope of the MS. For M $> 10^{11} $M$_{\odot}$, bulges are always charcaterised by red colours independent of their position on the LogSFR-LogM$_{\star}$ plane.

\begin{figure}
\centering
 \includegraphics[width=0.48\textwidth,keepaspectratio]{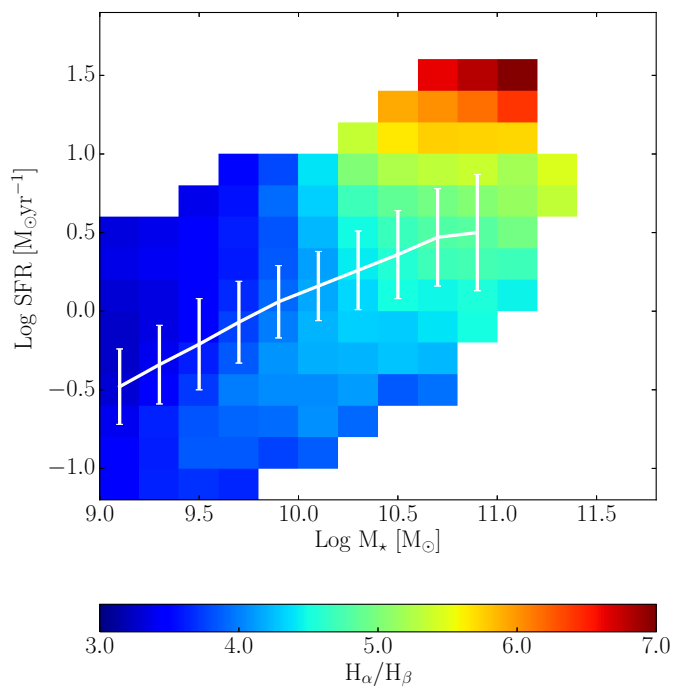}
\caption{Balmer decrement in the LogSFR-LogM$_{\star}$ plane for galaxies in our sample that have a median S/N ratio per pixel $>$ 8. The white solid line and error bars represent the MS of star forming galaxies and its dispersion.  }
\label{fig15}
 \end{figure}

In order to check the bias due to dust obscuration, we analyse the average Balmer decrement computed for galaxies with $\rm S/N > 8$, in both H$\alpha$ and H$\beta$ (Fig. \ref{fig15}). Dust obscuration is not significant for galaxies with stellar masses below $10^{10} $M$_{\odot}$. Hence the trends in colours reflect a trend in SF activity. Very massive ($\gtrsim 10^{10.5} $M$_{\odot}$) star forming galaxies, in the upper MS envelop, exhibit the largest values of Balmer decrement, pointing to a very high level of dust obscuration. This, in turn, would suggest that the flattening of the color gradients across the MS at increasing stellar mass in all panels of Fig. \ref{fig14} could be due to an increasing level of dust obscuration towards the upper envelope of the MS. 

We also point out that above $10^{10.5}$ M$_{\odot}$, the very red color of bulges in the green valley and on the MS can not be explained by a large level of dust obscuration due to the low average Balmer decrement. We conclude that the bulges in massive MS and green valley galaxies are intrinsically redder than their counterparts  in the passive region. This implies that such bulges might be older or more metal rich with respect to bulges in the passive region. In any case, this  suggests different evolutionary paths for bulges on the MS, and in the passive region. 

The observed trends of the bulge and disk colours at fixed stellar mass with the distance from the MS are preserved also when considering all galaxies in $S_{ALL}$ with 0.2$\le$B/T$\le$0.8 (a B/T range where galaxies have bulge+disk morphology).

 \section{Discussion and conclusions}
 \label{discussion}
 
We used the bulge-disc decomposition of Simard et al. (2011) to study the link between a galaxies' structural parameters and their location in the LogSFR-LogM$_{\star}$ plane. We use the colour of bulges and disks to understand how these individual components influence the evolution of the galaxy as a whole.  Our sample is drawn from the spectroscopic sample of SDSS DR7, in the redshift range $0.02<z<0.1$ and M$_{\star}>10^9$M$_{\odot}$, and the main findings are as follows:
 
\begin{itemize}
\item The MS of star forming galaxies is populated by galaxies that, at every stellar mass, have the lowest B/Ts. At low stellar masses, LogM$_{\star}<$10.2M$_{\odot}$, MS galaxies are pure discs, while for more massive galaxies the prominence of the bulge component increases with increasing stellar mass. 
\item The upper and lower envelopes of the MS are populated by galaxies characterised by intermediate B/Ts. This is robust against different decompositions, and independent of the modelling of the bar component;
\item Bulges in the upper envelop of the MS are characterised by blue colours at low stellar masses, or red colours and large dust obscuration at high stellar masses. This is consistent with high SF activity in the central region of the galaxy. 
\item The study of the mean bulge velocity dispersion and galaxy concentration parameter indicate that galaxies populating the upper and lower envelope of the MS are structurally similar. The values of the concentration parameter, in particular, suggest that these galaxies are characterised by classical bulges rather than pseudo-bulges.
\item In the low mass regime, M$_{\star}< 10^{10}$M$_{\odot}$, disc and bulge colours show a similar behaviour at fixed stellar mass, becoming progressively redder from the upper envelop of the MS to the passive region. Nevertheless, the reddening of the bulge component is steeper than for discs. For M$_{\star}> 10^{10}$M$_{\odot}$ disks and bulges become bluer when going form the MS towards its upper envelop, despite a less pronounced total variation of the colours. The trend of the bulge colour reverses in the lower envelop of the MS, where bulges are redder than in the passive region. 
\item The population of passive galaxies is largely made of genuine bulge + disc systems (at least 45$\%$).
\end{itemize}

Our results point to a tight link between the distribution of galaxies around the MS and their structural parameters. Contrary to Wuyts et al. (2011) and Cheung et al. (2012), we find that blue bulgy star forming galaxies are not outliers in the distribution of galaxies on the MS. They occupy the upper tail of the MS distribution leading to a progressive increase in B/T above the MS. In particular, Wuyts et al. (2011) find that the S\'ersic index remains $\sim 1$ in the region of the MS. We find, instead, that the MS location corresponds to the lowest value of the B/T at any mass. This discrepancy could be due to the inability of a single S\'ersic model to capture small bulges in disc dominated galaxies. 
In agreement with previous results, we find a large percentage of bulge dominated systems in the high mass SF region, where the MS scatter tends to increase \citep[i.e.][]{2014ApJ...788...11L,2014ApJ...795..104W,2016MNRAS.455.2839E}. 
However, the B/T-$\Delta_{MS}$ relation at fixed stellar mass holds in the stellar mass range $10^{9-11} $M$_{\odot}$, and seems to flatten where the MS itself tends to disappear. We also investigate how the presence of a bar influences the scatter around the MS. Overall we conclude that  bars do not affect the scatter of the MS or the B/T -$\Delta_{MS}$ relation. In fact, this relation holds for both the unbarred and the barred G09 subsamples. This suggests, partly in disagreement with the findings of \cite{Guo:2015kp}, that the predominance of the bulge in a galaxy is intrinsically related to the location of a galaxy around the MS, and this is true at any stellar mass independent of the presence of a bar.  Nevertheless, we do not exclude the possibility that a large fraction of bars among very massive SFGs can contribute to an increase in the scatter of the MS, as suggested by \cite{Guo:2015kp}.

\begin{figure}
 \includegraphics[width=0.95\columnwidth,trim={17cm 0cm 0cm 0cm},clip ]{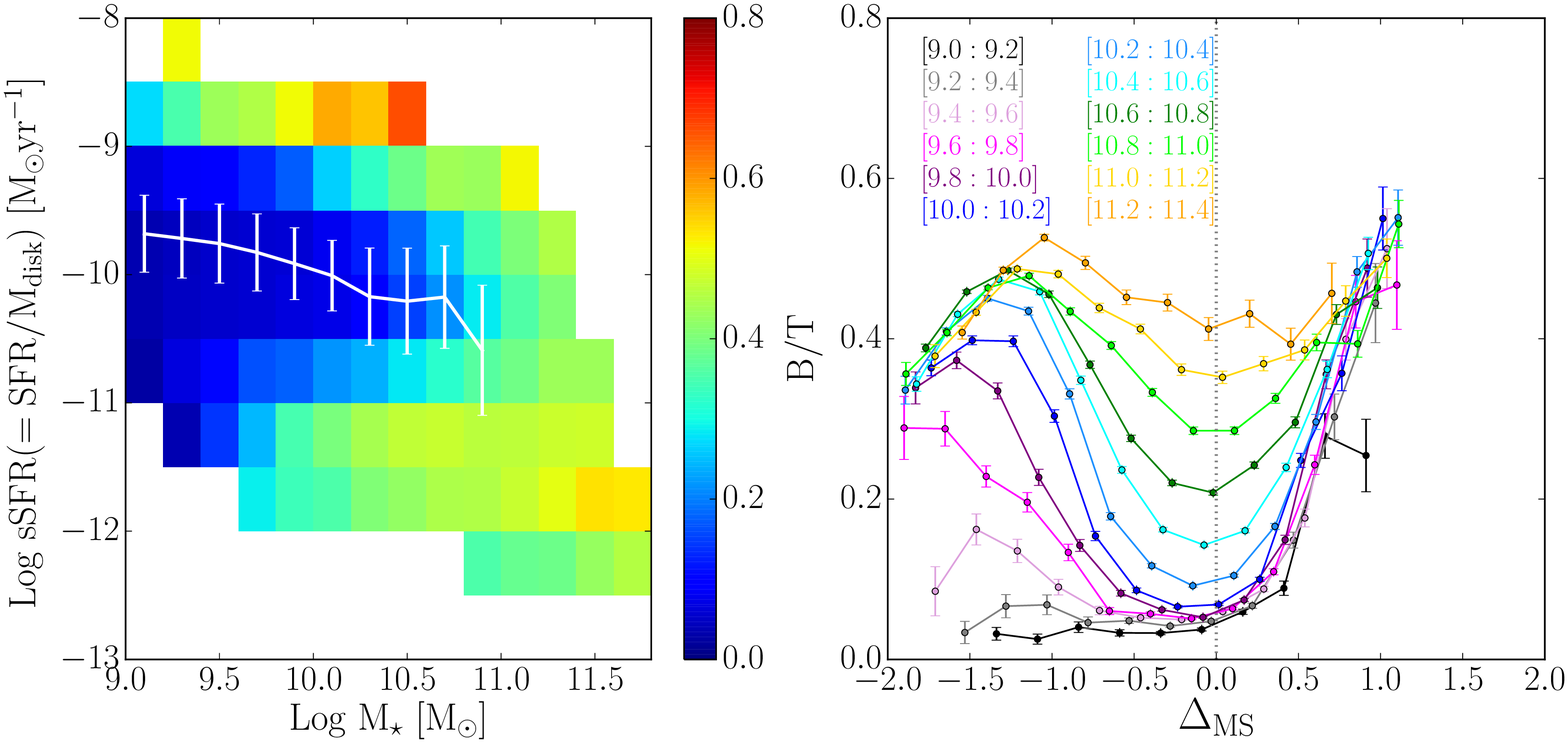}
\caption{B/T ratio as a function of the distance from the MS, represented by the vertical dotted line. Here $\Delta_{MS}$ has been computed from the sSFR normalised to M$_{disc}$. Different bins of stellar mass are indicated with different colours. }
\label{ssfr}
\end{figure}

Our study shows that the location of galaxies in the LogSFR-LogM$_{\star}$ plane and, in particular, with respect to the MS of SFGs, is determined by the combined activity (or inactivity) of the individual galaxy components. This was previously suggested by \cite{2014ApJ...785L..36A}, who pointed out that the MS becomes roughly linear if one assumes that the SF activity takes place only in the disc rather than in the entire galaxy. However, our results suggest otherwise. The effect of normalising the SF activity to the disc mass leads to a completely flat and tight relation only for pure disc galaxies, that, as shown here, dominate the core of the MS distribution. Extending this normalisation to galaxies of intermediate morphology has the additional effect of increasing the scatter by $\sim0.1$ dex in any mass bin. We suggest the following explanation:

\begin{equation}
 SFR_{gal}=SFR_{disc}+SFR_{bulge} 
\end{equation}
thus:
\begin{equation}
\frac{SFR_{gal}}{M_{disc}}=sSFR_{disc}+\frac{SFR_{bulge}}{M_{disc}}
\end{equation}

\noindent where  M$_{disc}$ is the disc mass and sSFR$_{disc}$ is the specific SFR of the disc. The last equation leads to the following effects. 

\begin{itemize}
\item[-]For pure disc galaxies, which dominate the core of the MS,  the mass of the disc equals the mass of the galaxy and SFR$_{bulge}\sim 0$, hence SFR$_{gal}/$M$_{disc}$ = sSFR$_{disc}$. Since the MS is nearly linear, sSFR$_{disc}$-M$_{\star}$ relation is flat with a very small scatter. 

\item[-]For intermediate B/T galaxies in the upper envelope of the MS, thus with a blue star forming bulge, SFR$_{bulge}> 0$ and M$_{disc}< $ M$_{star}$. This implies that SFR$_{gal}/$M$_{disc}$ = sSFR$_{disc}$ + SFR$_{bulge}/$M$_{disc}$. Therefore such galaxies are displaced well above the sSFR$_{disc}$-M$_{\star}$ relation by the contribution of SFR$_{bulge}/$M$_{disc}$. The larger the SFR of the bulge, the larger the displacement.

\item[-]For intermediate  B/T galaxies in the lower envelope of the MS, thus with a red quiescent bulge, SFR$_{bulge}\sim 0$, hence SFR$_{gal}/$M$_{disc}$ = sSFR$_{disc}$. They scatter around the sSFR$_{disc}-$M$_{\star}$ relation in the same way as around the MS.

\end{itemize}

In Fig. \ref{ssfr}, we show the relation between B/T and distance from the MS computed as the difference between the sSFR$_{disk}$ of galaxies and the sSFR$_{disk}$ of galaxies in the MS. This is done for galaxies that are pure disks, or have a secure bulge+disc structure. The B/T trends in Fig. \ref{ssfr} are similar to the ones in Fig. \ref{fig9}: a steep increase of the mean B/T in the upper envelop of the MS due to the displacement of bulgy SFGs above the sSFR$_{disc}-$M$_{\star}$ relation. We underline that the B/T in the passive region is lower that the one observed in Fig. 5, as here pure spheroidal galaxies are excluded from the sample. We find that the overall effect of neglecting the SF activity of the bulge component is to increase the scatter of the sSFR-M$_{\star}$ relation.


It has been proposed that minor mergers or violent disc instabilities could favour the flow of cold gas from the disc towards the galaxy centre and, thus, cause an overall compaction of the system. The high SF activity in the centre would lead to a compact, bulgy, star forming object \citep{2014MNRAS.438.1870D,2015MNRAS.450.2327Z}. \cite{2015Sci...348..314T} study the evolution of galaxies in this scenario using zoom-in simulations and they find that galaxies at high redshift undergo subsequent phases of compaction and depletion of the gas reservoir which ultimately leads to quenching. The compaction phase causes high SF in the central region of galaxies, and hence could favour bulge growth. This phase is then followed by depletion due to gas exhaustion. Such subsequent phases can move a galaxy across the MS: towards the upper envelope during compaction, and towards the lower envelope during depletion. Complete quiescence can be reached once the bulge reaches a given mass threshold corresponding to no more inflows in massive halos, or AGN feedback. Our findings on the B/T in the LogSFR-LogM$\star$ plane and bulge/disk colours can be related to such a scenario, and are also consistent with the observed gradient of molecular and atomic gas fraction across the MS, as seen by  \cite{2016MNRAS.462.1749S}.  



Integral Field Spectroscopy surveys like MaNGA \citep{2015ApJ...798....7B} and CALIFA \citep{2012A&A...538A...8S}  will greatly advance our understanding of the evolution of individual galaxy components, and how they impact the galaxy as a whole. This will help us in drawing a better picture of the quenching mechanism, and in understanding the CSFH at any epoch.

\section*{Acknowledgments}
We thank the anonymous referee for useful comments that improved the manuscript.  This research was supported by the DFG cluster of excellence ''Origin and Structure of the Universe''.  LM thanks Alvio Renzini and Pieter van Dokkum for their useful comments on the project. LM would also like to thank Bhaskar Agarwal for his inputs on the manuscript. 

\bibliographystyle{aa} 
\bibliography{laura} 

\begin{appendix}

\section{Results with S$_{ref}$ sample}
\label{Sref}

\begin{figure*}
 \centering
 \includegraphics[width=0.9\textwidth,keepaspectratio]{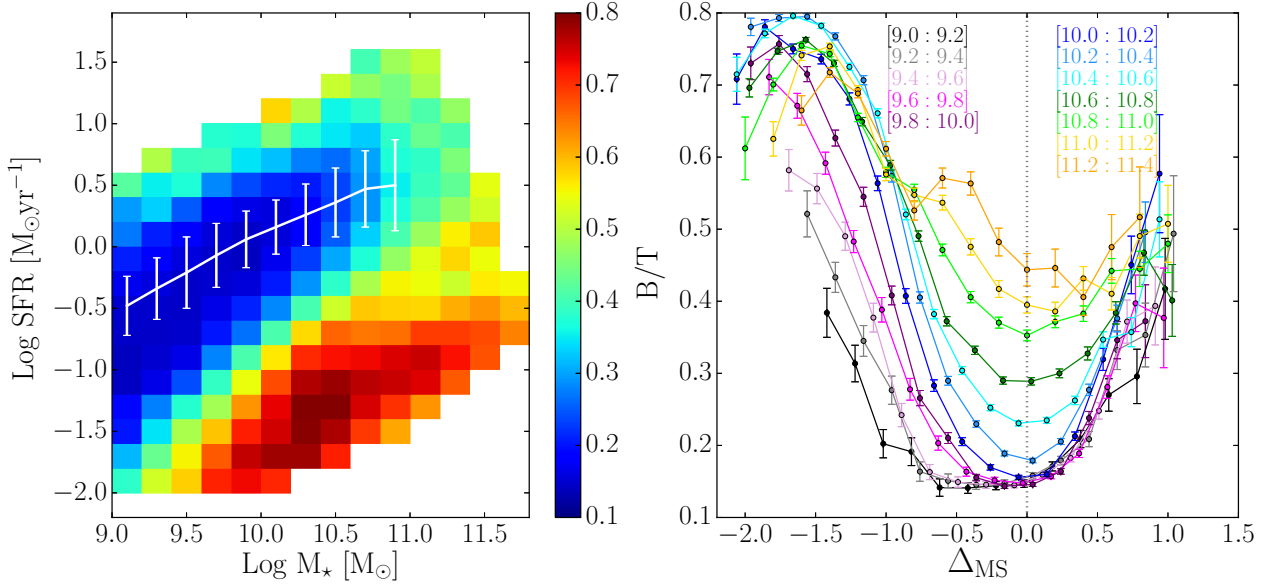}
\caption{Same as Fig. 5, but for S$_{ref}$ sample of galaxies. $Left\ panel$: LogSFR-LogM$_{\star}$ plane color coded according to the weighted average B/T in the bin. The white line represent the location of the MS of star forming galaxies, and the errorbars its dispersion. $Right\ panel$: B/T ratio as a function of $\Delta_{MS}$ (distance from the MS) in different stellar mass bins. }
\label{app.f1}
 \end{figure*}

Here we show the result of the analysis of the mean B/T in the LogSFR-LogM$_{\star}$ plane for the $S_{ref}$ sample. We recall that the $S_{ref}$ sample is built from the $S_{all}$ sample by keeping the B/T as given by S11 for all galaxies with $P_{PS}\le0.32$, while the remaining galaxies are considered as pure discs if they have B/T$\le$0.5, or pure spheroidals if they have B/T$>$0.5. We show the results in Fig. \ref{app.f1}. The trends of the B/T in the LogSFR-LogM$\star$ plane are remarkably consistent with those of Fig. 5 for the $S_{ALL}$ sample. We still observe the minimum of the B/T distribution for galaxies that, at each stellar mass bin, are located on the MS. As expected, the increase of the B/T in the passive region is more significant for $S_{ref}$ than for the $S_{ALL}$, as only the $\sim50\%$ of galaxies with B/T$>$0.5 has $P_{PS}\le$0.32. This confirms that our results are robust against the S11 decomposition.

\section{Concentration in the LogSFR-LogM$_{\star}$ plane: $R_{90}/R_{50}$}
 \label{append_comp}

\begin{figure*}
 \centering
  \includegraphics[width=0.9\textwidth, keepaspectratio]{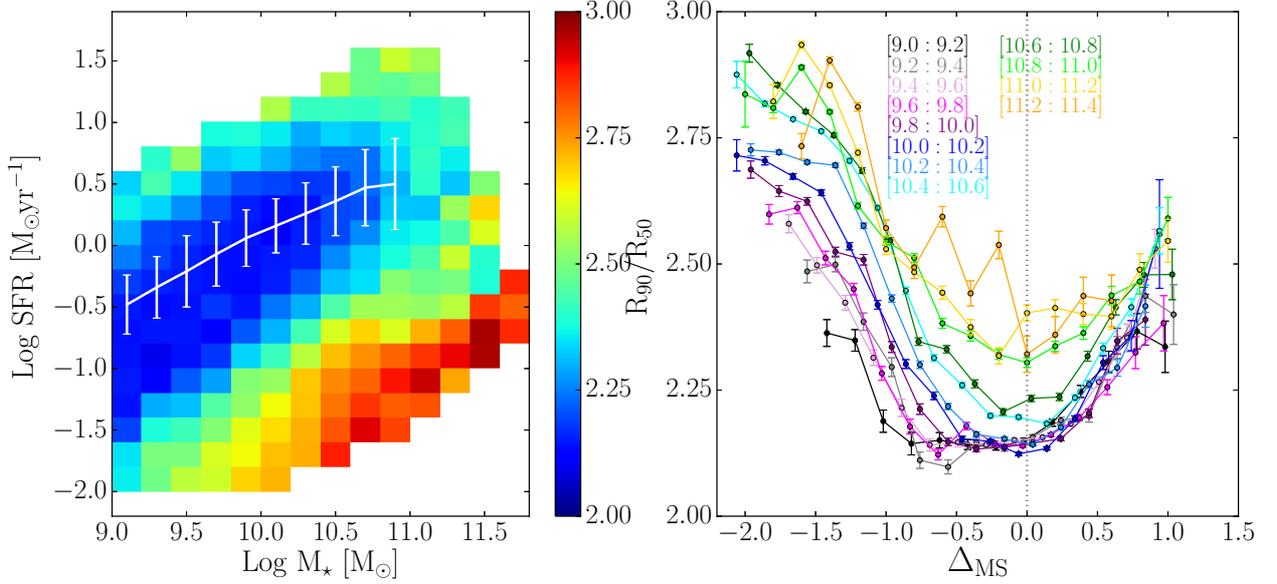}
\caption{$Left\ panel$: LogSFR-LogM$_{\star}$ plane color-coded as a function of the weighted average of the $R_{90}/R_{50}$ ratio. The white line represent the location of the MS of star forming galaxies, and the errorbars its dispersion. $Right\ panel$: $R_{90}/R_{50}$ ratio as a function of the distance from the MS ($\Delta_{MS}$) in different stellar mass bins.}
\label{fig.r90r50}
 \end{figure*}

We show here the behaviour of the concentration index ($R_{90}/R_{50}$) in the LogSFR-LogM$_{\star}$ plane in the $S_{ALL}$ sample. The values of $R_{50}$ and $R_{90}$ are taken from the SDSS DR7 catalogues. Fig. \ref{fig.r90r50} shows that the concentration index is minimum on the MS, and it increases both towards larger and smaller SFRs. Bulges in the upper and lower envelop of the MS are characterised by an average $R_{90}/R_{50}\approx$ 2.5 which, according to G09, is the typical concentration index of classical bulges with a low value of D4000, which suggest a young age (see Fig. 20 of G09). Classical bulges in the passive region have, on average, a $R_{90}/R_{50}$ of 3, typical of bulges with high value of D4000 suggesting an old stellar population. Pseudobulges are characterised by  $R_{90}/R_{50}\approx$ 2.0. We conclude that the upper and lower envelop of the MS are populated by bulgy galaxies with classical bulges with a relatively young population.

 \end{appendix}

\end{document}